# Magnetic Moment and Band Structure Analysis of Fe, Co, Ni-modified Graphene-nano- ribbon

Norio Ota

Graduate School of Pure and Applied Sciences, University of Tsukuba, *1-1-1 Tenoudai Tsukuba-city 305-8571, Japan*

Magnetic properties and band characteristics of graphene-nano-ribbon (GNR) modified by Fe, Co, and Ni were analyzed by the first principles DFT calculation. Typical unit cell is [$C_{32}H_2Fe_1$], [$C_{32}H_2Co_1$] and [$C_{32}H_2Ni_1$] respectively. The most stable spin state was $Sz$=4/2 for Fe-modified GNR, whereas $Sz$=3/2 for Co-case and $Sz$=2/2 for Ni-case. Atomic magnetic moment of Fe, Co and Ni were 3.63, 2.49 and 1.26 $\mu_B$, which were reduced values than that of atomic Hund-rule due to magnetic coupling with graphene network. There is a possibility for a ferromagnetic Fe, Co and Ni spin array through an interaction with carbon pi-conjugated spin system. By expanding a unit cell of Co-modified case as [$C_{96}H_6Co_3$], ferromagnetic like spin state and ferrimagnetic like one were compared. It was concluded that ferromagnetic state was more stable magnetic state. Band calculations of Co-modified case show half-metal like structure with relatively large band gap (0.55eV) for up-spin, whereas small gap (0.05eV) for down-spin. This suggested a capability of spintronics application like a spin fiter.

**Key words:** graphene, Fe, Co, Ni, spintronics, ferromagnetism, half-metal, density functional theory

## 1. Introduction

Current magnetic data storage[1)-2)] has a density around 1 tera-bit/inch$^2$ with 10 nm length, 25 nm width magnetic bit. We need a future ultra-high density spintronics material. One promising candidate is a molecule size ferromagnetic dot array having a typical areal bit size of 1 nm by 2.5 nm as illustrated in Fig.1. Recently, carbon based room-temperature ferromagnetic materials are experimentally reported[3)-9)]. They are graphite and graphene like materials. From a theoretical view point, Kusakabe and Maruyama[10)-11)] proposed an asymmetric graphene-ribbon model with two hydrogen modified (dihydrogenated) zigzag edge carbon showing ferromagnetic behavior. There are other important papers studied on graphene magnetism[28)29)]. Our previous papers[12)-15)] have reported multiple spin state analysis of graphene like molecules and nano-ribbons, which suggested a capability of strong magnetism. However, we need larger magnetization graphene-ribbon candidates.

Here, by the first principles density functional theory (DFT) based analysis, ferromagnetic atomic species like Fe, Co and Ni were tried to modify zigzag edge carbon of graphene-ribbon to find a capability of large magnetization. Typical unit cell were [$C_{32}H_2Fe_1$], [$C_{32}H_2Co_1$] and [$C_{32}H_2Ni_1$] respectively.

Already, there were some experiments of graphene formation on Fe (110) substrate[16)], and Co layer structure observation on graphite substrate[17)]. In DFT calculation, iron cluster on graphene sheet[18)], or iron –based molecule grafted on graphene[19)] were studied. Those were the cases that iron or cobalt atoms coupled with graphene surface sheet. In viewpoints of information storage and data processing devices, narrow stripe straight line like a graphene-nano-ribbon (GNR) is suitable for industrial application. Therefore, this report focuses on chemically modified GNR by magnetic species like Fe, Co and Ni.

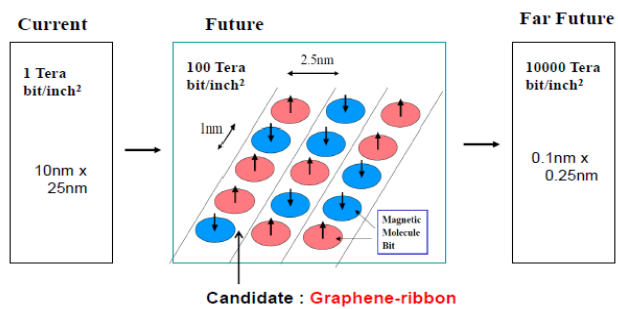

**Fig.1** Future ultra-high density 100 terabit/inch$^2$ magnetic data storage with a bit size of 1 nm by 2.5 nm. Graphene-nano-ribbon is one candidate.

## 2. Model graphene-nano-ribbon

Bird's eye view of typical graphene-ribbon model is shown in Fig.2. Iron atoms modify zigzag edge positioned carbons of one side (left hand side in the figure) of graphene-ribbon. Whereas, another side (right side) zigzag edges are all hydrogenated. As the first step of calculation, atoms were set initially as shown in Fig. 3(a). Isolated non-bonding Fe atom was positioned close to zigzag edge carbons. After repeating atom position optimization by DFT calculation, atomic configuration had converged as shown in Fig.3(b). One Fe-atom bonded with two zigzag edge carbons. Track width is 1.8 nm, tracking length of 1nm includes five zigzag carbon edges. Unit cell is shown in a square mark as [$C_{32}H_2Fe_1$], where blue ball show Fe, gray ball carbon and small ball hydrogen. In case of Co, unit cell is [$C_{32}H_2Co_1$] and Ni case [$C_{32}H_2Ni_1$].

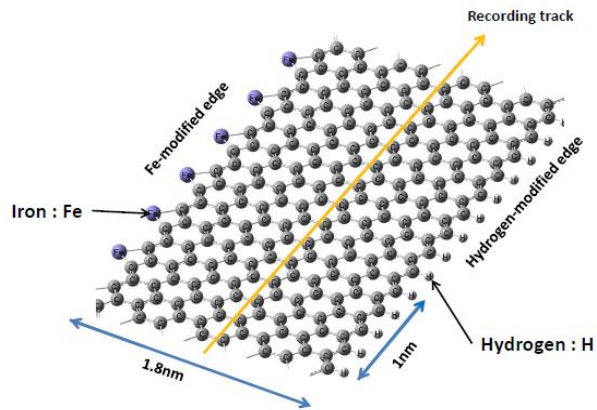

**Fig. 2** Bird's eye view of Fe-modified graphene ribbon. Iron atoms bond with zigzag edge carbons at the left side, whereas the right-side edge carbons are all hydrogenated.

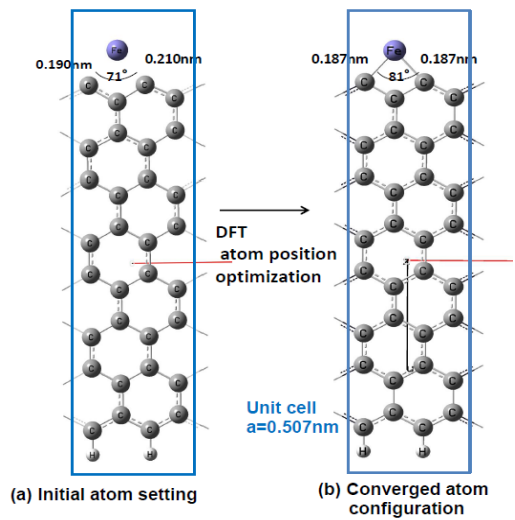

**Fig. 3** Initial atom setting is shown in (a). After atom position optimization, converged configuration is illustrated in (b). Unit cell is [$C_{32}H_2Fe_1$].

## 3. Calculation method

We have to obtain the (1) optimized atom configuration, (2) total energy, (3) spin density configuration, (4) magnetic moment of every atom and (5) band characteristics depending on a respective given spin state Sz to clarify magnetism. Density functional theory (DFT) [20)-21)] based generalized gradient approximation (GGA-PBEPBE) [22)] was applied utilizing Gaussian03 package[23)] with an atomic orbital 6-31G basis set[24)]. In this paper, total charge of model unit cell is set to be completely zero. In this unrestricted DFT calculation, S(S+1) value is obtained to check a degree of spin contamination[27)]. Inside of a unit cell, three dimensional DFT calculation was done. One dimensional periodic boundary condition was applied to realize an unlimited length graphene-ribbon. Self-consistent calculations are repeated until to meet convergence criteria. The required convergence on the root mean square density matrix was less than 10E-8 within 128 cycles.

## 4. Stable spin state

Typical GNR model is shown in Fig. 3(b). One unit cell has limited numbers of unpaired electrons, which enable allowable numbers of multiple spin states. In case of [$C_{32}H_2Fe_1$], there are five spin states like $Sz$=8/2, 6/2, 4/2, 2/2 and 0/2. Starting DFT calculation, one certain $Sz$ value should be installed as a spin parameter.

The question is which spin state is the most stable one. After spin parameter $Sz$ was installed, DFT calculation including atom position optimization was repeated until to meet convergence criteria. Converged energy for given $Sz$ was compared as shown in Table 1 in Rydberg-unit. The most lowest and stable spin state was marked by a large character in a blue box. In case of [$C_{32}H_2Fe_1$], the most stable spin state was $Sz$=4/2. This energy is close to that of $Sz$=2/2, therefore more precise energy calculation was carefully done again by applying more tight conversion criteria of density matrix less than 10E-10. Again, $Sz$=4/2 was the most stable one. Similar energy comparison was done for [$C_{32}H_2Co_1$] resulting the most stable one to be $Sz$=3/2, also for [$C_{32}H_2Ni_1$] to be $Sz$=2/2.

Additional important information was obtained from $S(S+1)$ value. In DFT calculation, we should mind spin-contamination[27)], which occur from unsuitable spin configuration. We can compare spin contamination by the difference between DFT obtained $S(S+1)$ value and installed $Sz(Sz+1)$ value. In case of $Sz$=4/2 of [$C_{32}H_2Fe_1$], $S(S+1)$ was 6.40, whereas $Sz(Sz+1)$ was 6.00, which were close together and suggested that this spin state is less spin-contamination one. Whereas, in case of $Sz$=2/2, $S(S+1)$ was 3.40 compared with $Sz(Sz+1)$ of 2.00, which means fairly large spin-contamination.

One typical spin density configuration was illustrated in Fig.4 in case of $Sz$=4/2 of [$C_{32}H_2Fe_1$]. Red cloud shows up-spin, while blue cloud down-spin. Contour lines for 0.001, 0.1, 0.4 e/A³ are predicted by arrows. Iron atom wears very large up-spin cloud. Bonded zigzag edge carbon show down-spin, whereas next neighbor carbon up-spin. Inside of graphene ribbon, up and down spins are orderly arrange one by one. This is a specific feature of pi-electron oriented spin configuration in graphene[15)]

Table 1 Comparison of calculated energy for selecting the most stable spin state. Blue box shows the most stable energy.

| Installed $S_z$ | DFT calculated energy (Rydberg) | | |
|---|---|---|---|
| | [C32H2Fe1] | [C32H2Co1] | [C32H2Ni1] |
| 0/2 | Not converged | -- | Not converged |
| 1/2 | -- | -2601.149716 | -- |
| 2/2 | -2482.122176 | -- | -2726.665682 |
| 3/2 | -- | -2601.149724 | -- |
| 4/2 | -2482.122180 | -- | -2726.634161 |
| 5/2 | -- | -2601.127446 | -- |
| 6/2 | -2482.096011 | -- | Not converged |
| 7/2 | -- | Not converged | -- |
| 8/2 | -2482.0342228 | -- | -- |

0.187nm, Co-C was 0.186nm for $S_z=3/2$ and Ni-C 0.185nm for $S_z=2/2$, which are reasonable with atomic radius. Angle of (C-Fe-C) was 81°, (C-Co-C) was 82°, and (C-Ni-C) was 82°. Distance of zigzag edge carbon to the nearest carbon was 0.145 nm, which is longer than graphene inside (C-C) distance of 0.141 nm.

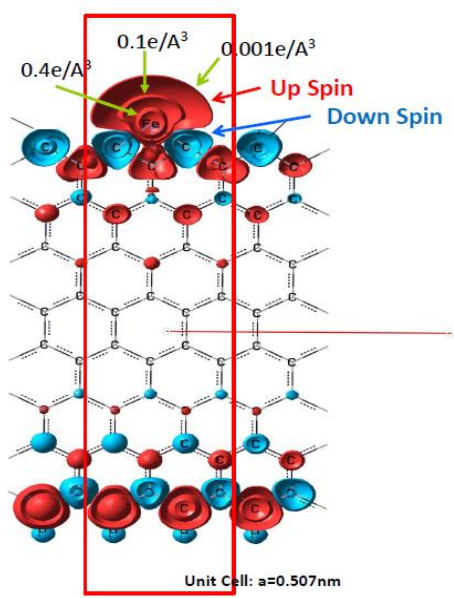

Fig.5 Spin density configuration in [$C_{32}H_2Fe_1$] GNR. Red cloud shows up-spin, and blue one down-spin. Fe-site has a very large up-spin cloud. Inside the graphene, up-spin and down-spin are orderly arrange one by one.

### 5. Optimized atomic configuration

Optimized atomic configuration show a flat and straight graphene ribbon. Some theoretical papers suggested that narrow width ribbon less than 1.5 nm may show some twisting and uniform curving[25)-26)]. However, our model has sufficient large width (1.8 nm) not occurring such irregularity. In Fig.6, calculated result of modified zigzag edge part was illustrated for (a) [$C_{32}H_2Fe_1$], (b) [$C_{32}H_2Co_1$], and (c) [$C_{32}H_2Ni_1$]. Distance between Fe-C (zigzag edge) for $S_z=4/2$ was

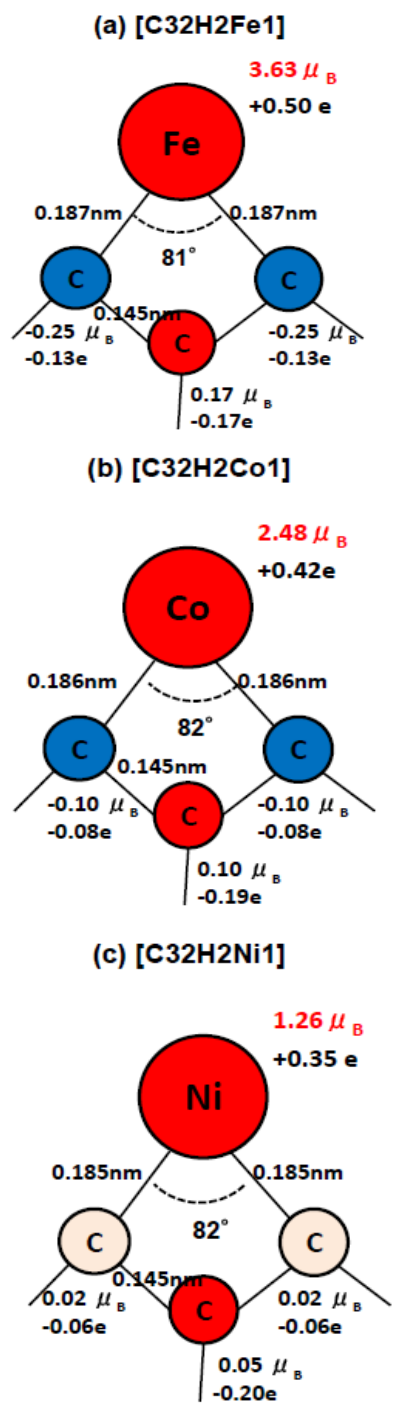

Fig.6 Calculated results of modified zigzag edge part of (a) [$C_{32}H_2Fe_1$], (b) [$C_{32}H_2Co_1$], and (c) [$C_{32}H_2Ni_1$]

## 6. Magnetic moment

Our purpose is to obtain a larger magnetization graphene-ribbon by Fe, Co and Ni modification. In case of stable spin state, DFT calculation gives an atomic magnetization $M$. In case of $[C_{32}H_2Fe_1]$, Fe-site has $M$(Fe)=3.63 $\mu_B$. In case of Co, $M$(Co) was 2.48$\mu_B$, and Ni case $M$(Ni)=1.26$\mu_B$. Those values were less than that of the Hund-rule[30] magnetic moments, 4.00$\mu_B$ for Fe, 3.00$\mu_B$ for Co, and 2.00$\mu_B$ for Ni. In Table 2, obtained charge and magnetic moment were summarized as (B), where isolated atom values were referenced as (A). In this table, (4s) means a sum of Gaussian-orbits[23] of 4s, 4px, 4py, 4pz, 5s, 5px, 5py, and 5pz. Also, (3d) is a sum of orbits of 6xx, 6yy, 6zz, 6xy, 6xz, 6yz, 7xx, 7yy, 7zz, 7xy, 7xz, and 7yz. Those many Gaussian type orbits represent natural occupied and unoccupied orbits.

In table 2, calculated charge of Fe is 7.50e, which is a sum of (4s) charge 1.26e and (3d) charge 6.24e. It should be noted that escaped charge from Fe(4s) (0.74e) partly donated to Fe(3d) (0.24e) and partly to other atoms (0.50e). Donated charge to Fe(3d) suggested a mixing of 4s and 3d electrons. Extra charge 0.50e was transferred to near three carbons with summed charge of -0.43e as shown in column (C) and Fig. 6(a). DFT calculated magnetic moment of Fe was $M$(Fe)=3.63$\mu_B$, which is a sum of $M$(4s)=0.37$\mu_B$ and $M$(3d)=3.26$\mu_B$. It should be noted that (4s) component has some magnetic moment with comparable order of near carbons shown in Fig.6. In case of Co and Ni, there were similar calculated results as presented in Table2.

Table 2 Charge and magnetic moment of Fe, Co, and Ni

| | | Isolated Atom (A) | DFT calculation results (B) | (A) – (B) | (C) |
|---|---|---|---|---|---|
| Fe of [C32H2Fe1] | Charge (e) | 8.0 | 7.50 | +0.50 | -0.43 e |
| | (4s) | (2.0) | (1.26) | (+0.74) | |
| | (3d) | (6.0) | (6.24) | (-0.24) | |
| | $M$ ($\mu_B$) | 4.0 | 3.63 | +0.37 | -0.33 $\mu_B$ |
| | (4s) | (0.0) | (0.37) | (-0.37) | |
| | (3d) | (4.0) | (3.26) | (+0.74) | |
| Co of [C32H2Co1] | Charge (e) | 9.0 | 8.58 | +0.42 | -0.35 e |
| | (4s) | (2.0) | (1.30) | (+0.70) | |
| | (3d) | (7.0) | (7.28) | (-0.28) | |
| | $M$ ($\mu_B$) | 3.0 | 2.48 | +0.52 | -0.10 $\mu_B$ |
| | (4s) | (0.0) | (0.39) | (-0.39) | |
| | (3d) | (3.0) | (2.09) | (+0.91) | |
| Ni of [C32H2Ni1] | Charge (e) | 10.0 | 9.65 | +0.35 | -0.32 e |
| | (4s) | (2.0) | (1.34) | (+0.66) | |
| | (3d) | (8.0) | (8.31) | (-0.31) | |
| | $M$ ($\mu_B$) | 2.0 | 1.26 | +0.74 | +0.09 $\mu_B$ |
| | (4s) | (0.0) | (0.26) | (-0.26) | |
| | (3d) | (2.0) | (1.00) | (+1.00) | |

$M$: Magnetic moment of Fe, Co, and Ni site
(c): Sum of neighbor three carbons' charge and magnetic moment

## 7. Possibility of ferromagnetic spin array

In order to check a possibility of ferromagnetic spin array, unit cell of Co- modified GNR was three times larger expanded as $[C_{96}H_6Co_3]$. Calculated spin configuration for $S_z$=9/2 was illustrated in Fig.7. Three cobalt sites wear all together up-spin cloud. This means a possibility of ferromagnetic spin order.

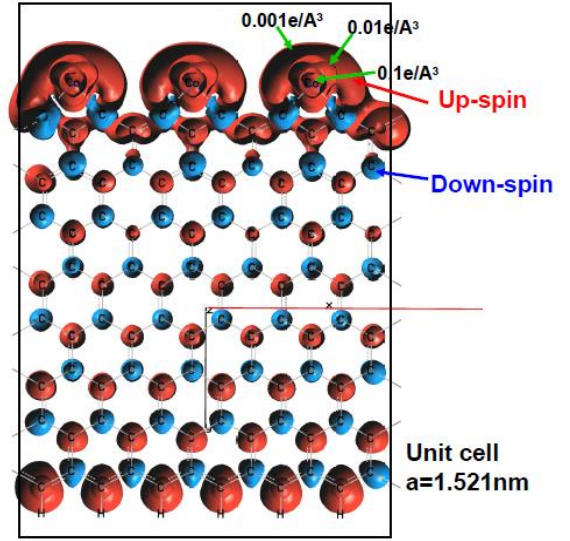

**Fig.7** Possibility of ferromagnetic spin order of Co was observed in three times expanded unit cell of Co-modified GNR $[C_{96}H_6Co_3]$. All cobalt sites have large up-spin cloud.

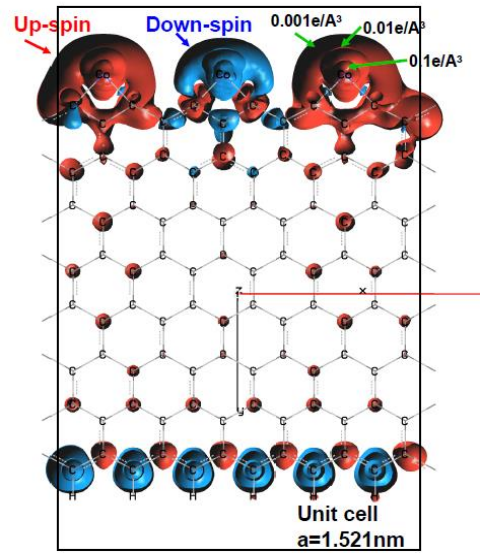

**Fig.8** Ferri-magnetic like Co-spin arrangement in case of $S_z$=3/2 in a unit cell $[C_{96}H_6Co_3]$. This state is unstable than ferromagnetic like $S_z$=9/2 state

There was no direct Co to Co exchange coupling in a resulted log file of DFT calculation. Whereas, including Co-spin, there are orderly arranged up-down spin pairs covering total inside of one unit cell. This is a similar spin configuration with graphene[15]. Co-atom was exchange coupled with zigzag edge carbon and played an up-spin part in graphene spin system, and demonstrated ferromagnetic like spin array.

On the other hand, ferrimagnetic spin configuration was obtained in case of $Sz=3/2$. As shown in Fig.8, there obtained Co-spin arrangement as like a ferrimagnetic like spin array of (up-down-up). It should be noted that there appear up-up spin pairs around Co and neighbor three carbons. Such up-up spin coupling increases the total energy. Obtained energy of ferrimagnetic like case of $Sz=3/2$ was -7804.388387 Rydberg. Whereas, in case of ferromagnetic like Sz=9/2 case, resulted energy was -7804.393449Rydberg. Ferromagnetic like state was lower energy level and suggested a stable magnetic state.

## 8. Band characteristics

GNR is a periodic system with one-dimensional crystallography. Here, Band characteristics were analyzed. In case of $[C_{32}H_2Co_1]$ unit cell, lattice parameter "a" is 0.507nm. We divided k-space to 12 elements from $k=0/a$ to $\pi/a$. In Table3, up-spin and down-spin energy gaps for every modified cases were compared. Up-spin gap was 0.55~0.64eV, whereas down-spin one was small value 0.05~0.19eV. Especially, in Co-case, detailed analysis shows interesting characteristics. Band structure of $[C_{32}H_2Co_1]$ is illustrated in Fig.9. Red curves are up-spin occupied orbits, light red unoccupied one, blue down-spin occupied and light blue unoccupied.

**Table 3** Band gap energy for Fe, Co, and Ni modified graphene-ribbon. Here, HOCO is the highest occupied crystal orbit, also LUCO is the lowest unoccupied crystal orbit.

|  | [C32H2Fe1] | [C32H2Co1] | [C32H2Ni1] |
|---|---|---|---|
| Up-spin HOCO (eV) | -3.99 | -4.04 | -4.16 |
| Up-spin LUCO (eV) | -3.36 | -3.49 | -3.51 |
| Up-spin gap (eV) | 0.63 | 0.55 | 0.64 |
| Down-spin HOCO (eV) | -3.67 | -3.83 | -4.27 |
| Down-spin LUCO (eV) | -3.54 | -3.77 | -4.08 |
| Down-spin gap (eV) | 0.12 | 0.05 | 0.19 |

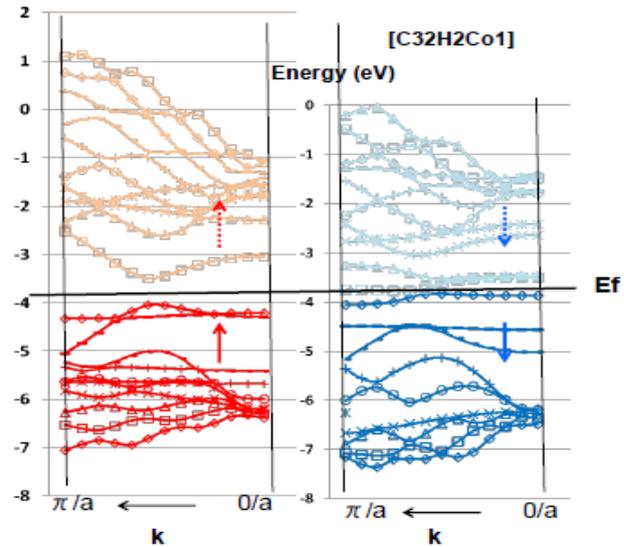

**Fig. 9** Band characteristics of $[C_{32}H_2Co_1]$.
Up-spin energy gap is 0.55eV. Among this gap, there are three down-spin orbits.

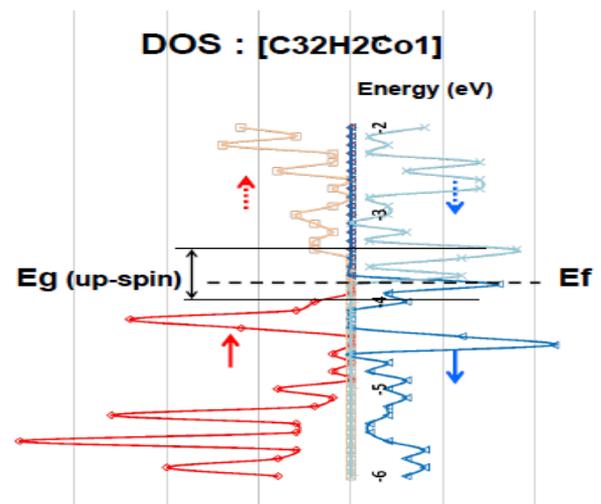

**Fig. 10** Density of states analysis of $[C_{32}H_2Co_1]$.
Up-spin energy gap Eg is 0.55e V, whereas down-spin states filled this gap, which suggests half-metal like characteristics.

As shown in Fig.9, up-spin energy gap is 0.55eV. Among this gap, there are three down-spin orbits (one occupied and two unoccupied). Density of state, as shown in Fig.10, suggested a down-spin dominated half metal like characteristics. In case of half metal, we can expect several spintronics devices as like a spin filter typically illustrated in Fig.11. Occupied up-spin orbits may cause resistive up-spin current. Whereas, down-spin orbits may give a conductive nature for down spin current. In an actual application, room-temperature operation is necessary. DFT calculation is essentially zero-temperature and ground state calculation method. However, if applying the Fermi distribution function modification, we can estimate an order of energy modification. Room-temperature 300K is related to energy of 0.025eV. Therefore, down-spin gap (0.05eV) has a possibility to be modified toward more small gap property suitable for half metallic devices.

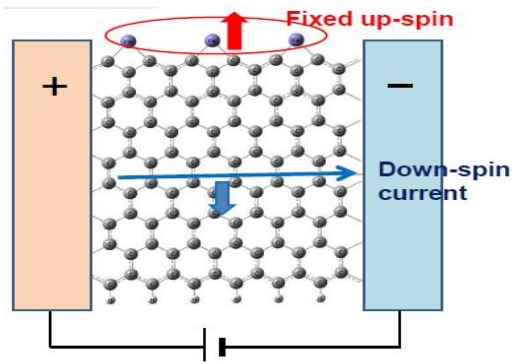

**Fig. 11** Spin filter device model using a half-metal like Co-modified GNR.

### 9. Conclusion

Magnetic graphene-nano-ribbon is very attractive candidate for achieving future ultra-high density spintronics devices. First principles analysis with density functional theory was carried out on Fe, Co, and Ni-modified zigzag edge graphene ribbon to increase its saturation magnetization. Typical unit cells were [$C_{32}H_2Fe_1$], [$C_{32}H_2Co_1$], and [$C_{32}H_2Ni_1$]. The most stable spin state was $S_z=4/2$ for Fe-modified graphene ribbon, whereas it was $S_z=3/2$ for Co and $S_z=2/2$ for Ni. The atomic magnetic moment of Fe was 3.63 $\mu_B$, that of Co was 2.49 $\mu_B$, and that of Ni was 1.26 $\mu_B$. These values were smaller than that of the atomic Hund-rule due to magnetic coupling with graphene network. There is a possibility for a ferromagnetic Fe, Co and Ni spin array through an interaction with carbon pi-conjugated spin system. In a Co-modified case, band calculations revealed a half-metal like structure with a fairly large band gap (0.55eV) for up-spin, whereas there was a very small gap (0.05eV) for down-spin, which will be useful features for spintronics devices such as a spin filter.

### Acknowledgment

I would like to say great thanks to Prof. Takeshi Inoshita of University of Tsukuba triggering this study.